\documentclass[twocolumn,aps,pre,showpacs,preprintnumbers,amsmath,amssymb]{revtex4-1}
\usepackage{graphicx}
\usepackage{color}
\usepackage{mathtools}
\usepackage{color}
\usepackage[normalem]{ulem}

\newcommand {\m}{{\kappa}}

\newcommand {\K} {\tilde K}
\newcommand {\w} {\omega}

\begin{document}
\title{ Phase coexistence  and spatial  correlations in reconstituting  $k$-mer models}
\author{Amit Kumar Chatterjee} 
 \email{amit.chatterjee@saha.ac.in} 
\author{Bijoy  Daga} \author{P. K. Mohanty}
\address{Condensed Matter Physics Division, Saha Institute of Nuclear Physics,1/AF Bidhan Nagar, Kolkata 700064, India}
\begin{abstract}
 In reconstituting  $k$-mer models, extended  objects which occupy  several   sites   on a  one dimensional lattice, 
 undergo  directed  or undirected diffusion,  and reconstitute  -when in contact-  by transferring  a single  monomer 
 unit from one  $k$-mer to the  other; the rates depend on the size 
 of  participating $k$-mers.  This  polydispersed system   has two 
 conserved quantities, the  number  of $k$-mers  and   the packing fraction. We 
 provide  a matrix product method to write the steady state of this  model and 
 to calculate the   spatial correlation functions analytically. 
 We show that   for a  constant reconstitution rate,  the  spatial  correlation
 exhibits damped oscillations in some density regions separated, from other  regions  
 with exponential decay, by a {\it disorder surface}.  In a specific limit, this constant-rate  reconstitution model 
 is equivalent to  a single dimer model and   exhibits  a phase coexistence  similar to 
the  one  observed earlier  in   totally asymmetric   simple exclusion process   
on a   ring  with a  defect.    
\end{abstract}
\maketitle
\section{Introduction}
Driven   diffusive  systems (DDS) evolve under local stochastic dynamics 
where by some conserved  quantity such as mass, energy or charge 
is being driven through the system  \cite{zia_book, dickman_book}.  
Compared to their equilibrium counterparts,
these  systems  exhibit  rich steady state behaviour \cite{krug_asep, derrida_mpa, evans_asep} including   
phase  separation \cite{phase_separation_1,phase_separation_2} and  condensation  transition \cite {evans_zrp} in one dimension,  
boundary layers \cite{B_layer}, localized shocks \cite{shocks}   and 
have  found a wide range of applications such as transport in super-ionic conductors \cite{KLS}, protein synthesis 
in prokaryotic cells \cite{gibbs_protein, zia_protein}, 
traffic flow\cite{traffic}, biophysical transport \cite{chou_bio_transport, frey_bio_transport} etc. . 
Recently, DDS with two or more  species   have been  studied \cite{multiSpcEP};  some of these  systems with more than one 
conserved quantities \cite{ABC_model, EKLS_model} also exhibit phase transition even in one dimension. It is argued  in \cite{phase_separation_2} 
that phase separation transition in DDS   is related  to 
the  condensation transition in a  corresponding zero-range process (ZRP ) \cite {evans_zrp}.

Driven  diffusive   systems  show interesting  steady state  behavior  when  the  constituting objects are   
extended \cite{tonks} in the sense that  they occupy   more  than one lattice site and move together as a single object  called $k$-mer,  obeying hardcore  constraints.
In one dimension, a driven system involving monodispersed $k$-mers was studied to understand the physical mechanism of protein synthesis in 
prokaryotic cells \cite{gibbs_protein}. For such a system, the time evolution of the 
conditional probabilities of the site occupation, starting from a known initial configuration has been calculated \cite{sasamoto}, 
and phase diagram for such systems has also been reported \cite{lee_protein}. Other works 
on such systems include studying hydrodynamic equations governing the local density evolution \cite{schutz1} and the effect
of inhomogeneities and defects \cite{kolomeisky_inhomogeneity, shaw_defect, dong_defect}. Microscopic processes like reconstitution, 
if present,   can  generate    $k$-mers  of arbitrary   lengths and 
facilitates the  possibility  of   phase separation.

In  a recent article \cite{daga_kmers} we have shown   that  diffusing and reconstituting   $k$-mers  can be mapped   
to  an interacting  box-particle system  with two  species  of particles.  This mapping helps us in finding the  exact  phase boundaries   of the phase separation transition   in   $k$-mer dynamics.   
Depending on   the   rate of the re-constitution dynamics, one  can  obtain a  macroscopic long polymer (which corresponds  to condensation of particles). At the same time,  since the   motion of  $k$-mers depends on their  size,
the system might go to a  phase  where   a large   $k$-mer   moves so  slow that  it    generates 
a large number of vacancies in front of it - this would lead to   condensation of the holes  ($0$s).   Of course,   in special situations,   one  may  encounter   simultaneous   condensation of   particles  and holes.  In  this  article, we aim at  calculating spatial correlation   in    reconstituting $k$-mer models. 
Spatial   correlation  functions, up to now,  has been calculated   for  models   with monodispersity, 
i.e. when all $k$-mers are of equal size \cite{Menon1997, zia_protein, gupta_kmer}.  It has been 
found that  steady state  of these  models   can be   written  in matrix product form 
and correlation function in these models  oscillate  in both  space and
time \cite{gupta_kmer}. Interestingly, in the continuum limit, the scaling behavior of 
the spatial correlation is found to be  same as obtained for a driven tonks gas 
\cite{tonks, tonks_correlation}. Various polydispersed models consisting of particles of 
different sizes and hence as many conservation laws have also been studied 
\cite{alcaraz_kmers, dhar_pico_ensembles,barma_kmers, grynberg_kmers}. 
Their phase behavior in general show strongly broken ergodicity and the dynamical critical 
behavior.

The    above   mentioned matrix formulation  for  fixed  size $k$-mers  can not   describe  the   polydispersed  systems 
where  the  $k$ mers  change their lengths dynamically.
In a mono-dispersed system, where  all  the $k$-mers are  of  equal  length, the  $k$-mer density automatically  
fixes the  packing fraction of  the lattice. In   reconstituting  $k$-mer   models, however,  the   density 
of $k$-mers   and  packing  fraction are  unrelated and  independently conserved. Thus, configurations  on a lattice, 
though contain   only $1$s  and $0$s, can  not  be  expressed as before   by   matrix  strings containing  just 
$D$s (for $1$s) and  $E$s (for $0$s) - an additional matrix must  be  introduced  to identify  each $k$-mer and to
keep track of   the conservation of $k$-mer density. 
It  turns out that  the additional  conservation  law  plays an 
important role in determining the stationary and dynamical properties of  reconstituting $k$-mer models. 
In this article  we  provide   a formalism    to write the stationary state of 
polydispersed $k$-mers in  matrix product from  and  calculate the spatial  correlation functions  analytically. 
We show   that 
when reconstitution occurs  between $k$-mers  of size  $k_1$ and $k_2$  with rate  $w_1(k_1) w_2(k_2),$ 
one can {\it always}   write   an   infinite dimensional representation  of matrices  in terms of the rate functions.
However, some specific   cases  can be   represented   by  finite dimensional matrices. One such example  is  the 
constant-rate   reconstitution (CRR) model 
where reconstitution occurs with constant rate and monomers  diffuse with a rate different from the  the other $k$-mers.
We calculate  spatial  correlation functions of  CRR model explicitly  and find that they  show  damped oscillations  
in  some parameter  regime and  decay exponentially  in other  regimes. The  disorder line that separates  these  regimes 
is also calculated.

The     article  is organized as follows. In section \ref{sec:model}   we introduce  
the  reconstituting  $k$-mer models    and  develop the 
matrix formulation  to  calculate    the steady state   weights  of  configurations from  representing  string of matrices. 
In section   \ref{sec:example},  we introduce   the CRR  model,   which  has a  finite  dimensional matrix representation, 
and calculate   the  spatial correlation functions  and the disorder line  explicitly  for  a   given diffusion rate.
Also, in this section   we show that   the CRR model in a special  limit, exhibits  phase coexistence   similar to the  one observed in  asymmetric exclusion process   with  a  single  defect.    Finally,  we   conclude  and  summarize the results in section   \ref{sec:conclude}.

\section{The  reconstituting $k$-mer  model \label{sec:model}} 
Let  us   consider a  driven diffusive system  of   polydispersed $k$-mers  on a one  dimensional periodic lattice 
involving the directed diffusion and  reconstitution  dynamics.  Along with drift,  the 
$k$-mers   change their    size  through exchange of monomer units. It is assumed that  the reconstitution  dynamics  
does not allow  complete fusion of    monomers  and thus, not only the   mass (the  total length of the  $k$-mers) but also 
the number of  $k$-mers     is   conserved.  

\begin{figure}[h]
 \centering \includegraphics[width=8cm]{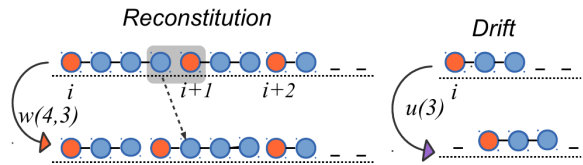}
\caption{(Color online) Poly-dispersed $k$-mers in  one dimension showing drift and reconstitution. 
Engine  of the   $k$-mers  are  marked  as orange.  The rate of drift $u(k)$ depends on the length 
of the corresponding $k$-mers.  Reconstitution occurs  only  among  consecutive   immobile $k$-mers
with a  rate $w(k_i,k_{i+1})$  which depends on their lengths. An additional constraint $w(k,k') = 0$ for $k<2$, conserves the 
number of  $k$-mers in the system.
}
\label{fig:dyn}
\end{figure}
For completeness, we start with the  model   studied  recently in \cite{daga_kmers}.
Let us  consider $M$ number of  $k$-mers   on   on a one dimensional  periodic lattice of
$L$ sites labeled by the index $i = 1, 2, . . . ,L.$  The  $k$-mers,  each    having  different   integer 
  length,  are also  labeled  sequentially as  $m=1,2,\dots,M.$

A $k$-mer is a hard extended object  which occupies $k$  consecutive  sites on a lattice, and can be  denoted by 
a string of  $k$  consecutive $1$s
(here represented by $1^k$). 
Thus, every configuration of the system can be 
represented by  a binary sequence  with each  site $i$  carrying a variable  $s_i= 1$ or $0$  
denoting  respectively whether the  lattice site is occupied  by a $k$-mer or not.
The total number of vacancies ($0$s) in the system is $N$ and thus, the  total 
length of the  $k$-mers  is  $K= \sum_{m=1}^M k_m = L-N$  (total number of $1$s).
We  define the free volume (or void density) as $\rho_0=N/L$  and the  $k$-mer  number density 
as  $\rho=\frac{M}{L}$.  Thus  the packing fraction   of the lattice (fraction of volume  occupied by the  $k$-mers)  
is $\eta=1-\rho_0=K/L.$   

We consider  directed diffusion of  $k$-mers;  a  $k$-mer   of size $k_m$  moves to its right with rate  $u(k_{m}),$
\begin{equation}
 k_m 0    \xrightarrow{u(k_{m})}  0k_m     ~~\equiv~~   1^{k_m}0   \xrightarrow{u(k_{m})}  0 1^{k_m}.
 \label{eq:dyn1}
\end{equation}
Along with  this,   reconstitution  occurs   among neighboring   $k$-mers 
where  one of the   $k$-mers may release  a single unit (or monomer)  which instantly joins the other $k$-mer (see  Fig. \ref{fig:dyn}).
Note  that, reconstitution  dynamics acts only   at the interface of  {\it immobile}   $k$-mers which   are expected to  remain 
in contact for long, 
\begin{equation}
(k_m,~k_{m+1}) \xrightleftharpoons[w(k_{m+1}+1, k_m-1)]{\,w(k_m,k_{m+1})\,} (k_m-1,k_{m+1}+1).
\label{eq:dyn2}
\end{equation}
 The reconstitution rate $w(k_m,k_{m+1})$ depends on  the  size of the  participating $k$-mers and constrained by 
a condition $w(1,y) =0,$   which  prohibits merging of $k$-mers and     maintains   the conservation  of 
$k$-mer   density $\rho.$ It is  evident that  Eqs.   \eqref{eq:dyn1} and  
\eqref{eq:dyn2} also  conserve  $\rho_0.$

The dynamics of the model can be mapped  exactly to a  two-species  generalization of  misanthrope process  (TMAP)
\cite{footnote_TMAP}
 by considering  each $k$-mer  as a box containing  $\m= k-1$ particles  of one kind ($k$-particles)
 and  the number of consecutive vacancies (say $n$)  in front   of  the $k$-mer   as the  number  
 of particles of other type ($0$-particles). Thus,  in TMAP,  we have $M$ boxes containing  
 $\tilde K= K-M$  number of  $k$-particles and $N$  number of $0$-particles.

From a box $m$  containing  $(n_m, \m_m)$  number of   $0$- and $k$-particles respectively,  the particles hop 
to one of the the neighboring boxes $m'= m\pm1$ having  $(n_{m'}, \m_{m'})$ particles  with rates $u_0$ and $u_k$ (respectively for 
$0$- and $k$-particles),
\begin{eqnarray}
  && u_0(\m_m) =  u(\m_m+1)  \delta_{m', m-1}\cr
 &&u_k(\m_m, \m_{m'}) 
 =w(\m_m+1, \m_{m'}+1) \delta_{n_m,0}   \delta_{n_{m'},0}.\label{eq:product}
\end{eqnarray}
 The $\delta$-function  in the  first equation  forces     $0$-particles to move towards  left 
 (same as  $k$-mers  moving to the right  neighbor on the lattice) and those in the  last equation  take  care of  
 the  restrictions  that  reconstitution   (or exchange  of $k$-particles) occurs among  boxes  $m$ and $m'$  only  
 when   they  are  devoid of $0$-particles (equivalently, when  $k$-mers are immobile).

It is well known that misanthrope  processes  enjoy the   luxury of  factorized steady  
state \cite{EvansBeyondZRP} for  hop rates satisfying certain specific conditions.  It  is straight forward 
to  derive similar conditions on hop rates  of TMAP
so  that  its steady state   has a   factorized form, 
\begin{equation}
    P({\m_i,n_i}) =  \frac{1}{Q_{\K,N}} \prod_{i=1}^{M} f(\m_i,n_i) \delta
   \small { \left(\sum_i  \m_i -\K \right)\delta \left(\sum_i  n_i -N\right)}, \label{eq:FSS}
\end{equation} 
where the  $\delta$ functions ensure   conservation of  $\K,N,$  the total  number of  particles of each species,  
and $Q_{\K,N}$ is   the  canonical partition  function.
When  the  reconstitution   rate  in $k$-mer  model  has a  product form  $w(k,k')=w_1(k)w_2(k'),$   
the  hop rate   of   $k$-particles  
in corresponding  TMAP  (see  Eq. (\ref{eq:product}))  also takes a product form $u_k(\m,\m') = w_1(\m)w_2(\m').$ 
For  this  simple choice,  the weight function  is  
given by \cite{daga_kmers}, 
\begin{equation}
f(\m,n)=\frac{1}{[u_0(\m)]^n} \prod_{\m'=1}^{\m} \frac{w_2(\m'-1)}{w_1(\m')}. \label{eq:fkn}
\end{equation}
Once  the functional forms of  $u(.), w_1(.), w_2(.)$   are specified,   one can calculate the steady state 
properties  of the  TMAP exactly from  the partition function  in  grand canonical ensemble (GCE) 
using two fugacities $x$ and $z$ for  the conservation of   $\K$ and $N$ respectively,
\begin{equation}
Z(x,z) = \sum_{\K=0}^\infty \sum_{N=0}^\infty Q_{\K,N} x^{\K} z^N .
\end{equation}
From this   partition   function,  one can  further   calculate   \textit{one-point   functions}   
of  the  $k$-mer models   analytically. However,  spatial  correlation  
functions   can not be calculated  straightforwardly.
This  is because   the site variables  $s_i=1,0$ on the  $k$-mer model, which  represent whether the site is occupied by 
 a $k$-mer or not, are   not so simple functions  of the occupation numbers  $\{ n_i, \m_i\}$ of  TMAP. We  have,   
 \begin{equation}
  s_i =  1-  \sum_{j=1}^M \theta(g_j-i) \theta(i+2+n_j -g_j); g_j= \sum_{l=1}^j  (1+  n_l + \m_l);
 \end{equation}
where  $\theta(x)$ is the Heaviside theta  function.  Clearly   obtaining spatial correlation functions 
$C(r) = \langle  s_i s_{i+r}\rangle -\langle  s_i \rangle^2$ would be   difficult (though not impossible)  from  the  TMAP correspondence. 
In the  following, we  provide  a  matrix   formulation  to  obtain  the  steady state  weight  of any 
configuration   of  the $k$-mer model from a  matrix  string  which  uniquely  represents that configuration.  

\subsection{Matrix product steady state}

To calculate the  spatial correlation functions  explicitly and conveniently,  in this section, 
we   provide  a  matrix formulation  similar to the  one  obtained  earlier
for exclusion processes  having ZRP  correspondence \cite{OZRP}.  
 In \cite{OZRP}, the authors showed that steady states of 
 one dimensional exclusion models  having  ZRP correspondence can be written in matrix product 
 form -they also provided an infinite dimensional representation of the matrices that can always be obtained 
 from the corresponding ZRP weights \cite{footnote3}.   In this article we try to obtain  the spatial correlation 
 functions of a polydispersed  systems   in a similar way, i.e.,  by writing the  steady state weight  of 
 configuration  as  the trace  of  a representative matrix string. 
 
 How many matrices  do we need ? In ZRP,  or its equivalent  exclusion process,  there  was only  one 
 conserved quantity, which is the  density  or the packing fraction $\eta.$  Here, we have an additional 
 and   independent conserved quantity $\rho,$ the $k$-mer density.
 Thus  along with  matrices $D$ and $E$  which represent the occupation status of a 
 site we  need another matrix, say $A$, that  would appear once for every $k$-mer so that the $k$-mer number density is fixed
 appropriately.  It is convenient to assign  this matrix $A$ to the  left most site (or an engine) of  
 a $k$-mer.   In other words, every $k$-mer  ($1^k$ on a  lattice) is represented  by  $A D^{k-1}.$    
 In   summary, in the   matrix formulation, all occupied sites except the engine $(A)$, are   represented by   matrix $D$s,    
 and  the   vacant sites  are  represented by  $E$s.  Now, the  steady state   weight  of a configuration   $\{n_i,k_i\}$ 
 can be expressed  (using  $\m_i= k_i-1$   in  Eq. (\ref{eq:FSS}))    as 
  \begin{equation}
  \prod_{i=1}^{M}  f(k_i-1,n_i) = Tr \left[ \prod_{i=1}^{M}  A D^{k_i-1} E^{n_i} \right]. \label{eq:fkn1}
  \end{equation}
  We further assume that   $A$   can be  expressed as an  outer product of two vectors
  $A=|\alpha \rangle \langle \beta|$;   the  vectors  $|\alpha\rangle,$ $\langle \beta|$  and matrices 
  $D$ and $E$    need to be determined from the  dynamics.   With this choice,  Eq. (\ref{eq:fkn1})   results in, 
 \begin{equation}
f(k,n)=\langle \beta|D^k E^n|\alpha \rangle. \label{eq:main}
\end{equation}
This equation is generic  as  long as the  steady state   of the   TMAP corresponding to  a reconstituting $k$-mer model has  
a  factorized steady state. Now, any representation of $A=|\alpha \rangle\langle \beta|,D,E$ that   satisfy 
Eq. (\ref{eq:main})  can   provide   a  matrix  product steady state   for  the   $k$-mer model.  
One must, however, remember that  any  arbitrary matrix string  does not  necessarily represent a  configuration of 
the  $k$-mer model. Since   every block of vacant sites (string of  $E$s)  must   end  with  a $k$-mer  represented by $AD^{k-1},$ 
all valid  matrix  strings  must be  devoid of $ED.$ This   brings in an additional constraint, 
 \begin{equation}
ED=0 \label{eq:de0}
\end{equation}
which must be  accounted  for while searching  a  suitable matrix representation. 
 
 We  now restrict ourselves to specific $k$-mer dynamics   which   leads to a   factorized steady state,  as   in  Eq. (\ref{eq:FSS}).
 For the $k$-mer  with hop rate $u(k)$  and reconstitution rate $\w(k,k')=w_1(k)w_2(k')$,  the steady state is   given by 
 Eq. (\ref{eq:fkn}).  A  set of matrices   $\{ \tilde A,\tilde D, \tilde E \}$   that  satisfy  Eq. (\ref{eq:main}) 
  \begin{equation}
 \langle \tilde \beta|\tilde D^k \tilde E^n|\tilde \alpha \rangle=\frac{1}{[u(k+1)]^n} \prod_{k'=1}^{k} \frac{w_2(k'-1)}{w_1(k')}
 \end{equation}
is   
 \begin{eqnarray}
 |\tilde \alpha \rangle= \sum_{i=1}^\infty |i \rangle; \langle \tilde \beta| = \langle 1|;  
 \tilde D= \sum_{i=1}^\infty \frac{w_2(i-1)}{w_1(i)}  |i \rangle \langle i+1| \cr
 \tilde E= |1 \rangle \langle 1|  + \sum_{i=2}^\infty \frac{1}{u(i-1)} |i \rangle \langle i| \; \;\;\;\;\;\; \;\;\;\;\;\; \;\;\;\;\; 
 \label{eq:tild}
 \end{eqnarray}
 where $\{|i \rangle\}$     with $i=1,2\dots$    are  standard basis  vectors  in    infinite dimension. 
 These matrices,   however, do not satisfy Eq. (\ref{eq:de0}). One option is to  discard  this  representation 
 and look for a  new one that   satisfy both  Eqs. (\ref{eq:de0}) and  (\ref{eq:main}),  which can be done 
 in   certain  specific  cases (see next section).  But Eq.  (\ref{eq:tild}) is a general representation  
 for  $k$-mer models where   $k$-mers   drift  with rate  $u(k)$ and  reconstitute
 with   a  rate having product form $w(k,k')= w_1(k) w_2(k')$. Thus it would be beneficial to hold on to
 these matrices  $\{  \tilde A, \tilde D,\tilde E \},$ and to construct a  new representation   
 using  them, which    satisfy   both Eqs. (\ref{eq:main}) and (\ref{eq:de0}). In this context, the following 
 representation works:
 \begin{eqnarray}
 |\alpha \rangle=  |\tilde \alpha \rangle \otimes (|\tilde 1 \rangle    + |\tilde 2 \rangle), ~~
  \langle  \beta| = \langle \tilde \beta| \otimes \langle \tilde 2|, \cr
  D=\tilde D  \otimes |\tilde 2 \rangle\langle \tilde 2|, ~~ 
  E = \tilde E \otimes (|\tilde 1  \rangle  + |\tilde 2  \rangle) \langle \tilde 1|.
  \label {eq:cross}
 \end{eqnarray}
 Here, $\{ |\tilde 1 \rangle, |\tilde 2 \rangle \}$ 
are the standard  (and complete)  basis  vectors  in  $2$-dimension.  More explicitly, we have, 
\begin{equation}
\begin{array}{ccc}
A=\left( \begin{array}{cc}
0 & \tilde A \\
0 & \tilde A \\
\end{array}
\right), &
D=\left( \begin{array}{cc}
0 & 0 \\
0 & \tilde D \\
\end{array}
\right), & E=\left( \begin{array}{cc}
\tilde E & 0 \\
\tilde  E & 0 \\
\end{array}
\right) \\
\end{array}.\label {eq:cross_mat}
\end{equation}
This  infinite dimensional representation provides  a  matrix  product steady state   for drifting and
reconstituting $k$-mers in one dimension as  long as  the  re-constitution rate has a product form.     
In the following, we illustrate  a specific case where the representation is  finite dimensional.

\section{Constant-rate  reconstitution (CRR)  model \label{sec:example}}
In this section we  study a   specific example  of   reconstituting $k$-mer   model  and illustrate the 
matrix  product formulation  presented  in previous section.   Let us consider  that  the $k$-mers  drift 
to their right neighbor   with rate $u(k) = 1 + (v-1) \delta_{k,1},$  i.e.,  the monomers ($k=1$) move with rate   
$v$ whereas other $k$-mers move  with  unit rate.  Let the   re-constitution  rate  be  a constant $\w,$
independent of  the size of the  $k$-mers.  In this  constant-rate  reconstitution (CRR) model matrices 
$\{\tilde A, \tilde D, \tilde E \},$   which satisfy only  Eq. (\ref{eq:main}),
have   two dimensional representations, 
\begin{equation}
\begin{array}{ccc}
\tilde A=\left( \begin{array}{cc}
1 & 0 \\
1 & 0 \\
\end{array}
\right), &
\tilde D=\w \left( \begin{array}{cc}
0 & 1 \\
0 & 1 \\
\end{array}
\right), & \tilde E=\left( \begin{array}{cc}
\frac{1}{v} & 0 \\
0 & 1 \\
\end{array}
\right) \\
\end{array}.
\end{equation} 
However,  since these matrices   do  not   satisfy   Eq. (\ref{eq:de0}), 
we now  construct  new  matrices   $\{ A,D, E \},$ using  Eq.  (\ref{eq:cross}), 
or  (\ref{eq:cross_mat}),
\begin{eqnarray}  
A= \sum_{i=1}^4 |i\rangle  \langle  3|,\: \: D= \w \sum_{i=3}^4 |i\rangle  \langle  4|,\; \;\;\;\; \cr 
E = \sum_{i=1}^2 ( \frac{1}{v}  |2i-1\rangle  \langle 2i-1| + |2i\rangle  \langle  2i|) 
\label{eq:rep_v2}
  \end{eqnarray}
which    are   4-dimensional, and  represent   respectively  the  engine  of  the $k$-mer,  any other unit  of the  $k$-mer 
and the  vacancies  $0$s.

To calculate the correlation function   and the densities, as   usual,  we  start with    the   partition function in GCE, 
$Z = Tr[T^L]$  where   the  transfer matrix   $T= A + x D + z E .$   The  fugacities  $z$ and  $x$ together control  the   
densities $\rho_0$   and  $\rho_d= 1-\rho-\rho_0$, representing density of  $E$s  and  $D$s respectively. 
In fact, in  this problem,  the    transfer matrix  $T \equiv T(z, x \w, v)$    does  not depend  independently    on 
$x$ and $\w$,  rather   depends on their product. Thus,  any  particular  value   of $\w$ only redefines the 
fugacity   $x\to x \w$  and  we can  set $\w=1$ without loss of generality.  Now, the 
partition function in GCE :
 \begin{equation}
 Z(z,x)= Tr (T^L)  ~;~ T =A + x D + z E   =
 \left( \begin{array}{cc}
z \tilde E &  \tilde A \\
z \tilde E &   x \tilde D+ \tilde A \\
\end{array}\right).
\end{equation}
The   characteristic equation    for  the eigenvalue of $T$ is $\lambda f(\lambda)=0,$ with 
\begin{eqnarray}
  f(\lambda) &=& x z (1 - v -z)  + \{v+ z + x(1+ v) \}  z \lambda \cr && ~~~~ - \{z+ v (1+x+z)\} \lambda^2 + 
  v \lambda^3. \label{eq:char} 
\end{eqnarray}
 Thus  one of the   eigenvalues of $T$ is  $0$ and   other  three, denoted  by  the largest eigenvalue $\lambda_{max}$,  
 and  $\lambda_{1,2}$,  are   roots   of the cubic polynomial $f(\lambda).$
The   partition function in GCE  is then, 
 \begin{eqnarray}
   Z(z,x)&=&  \lambda_{max}(z,x) ^L   +  \lambda_1(z,x) ^L +   \lambda_2(z,x) ^L  \cr &\simeq&  \lambda_{max}(z,x) ^L
 \end{eqnarray}
where in the last step we have taken  the thermodynamic limit  $L\to \infty.$   The conserved densities of the   
canonical ensemble   are now, 
 \begin{equation}
  \rho_d = x \frac{\partial}{\partial x} \ln  \lambda_{max},
 \,\, \rho_0 = z \frac{\partial}{\partial z} \ln  \lambda_{max}.
 \label{eq:dens-fug}
 \end{equation}
This  density fugacity  relation   specify  the  values of $(z,x)$  which   uniquely correspond  to  a  particular   
pair of conserved densities $( \rho_0,\rho_d).$  Any   other observable  in GCE, which are  functions of   $(z,x),$    
can be  expressed in terms of   the  densities  $( \rho_0,\rho_d)$  using  (\ref{eq:dens-fug}).  We  must mention that 
densities can also be obtained as follows.
 \begin{eqnarray}
  \rho_d = \langle  d_i \rangle&=& x \frac{ Tr [ D T^{L-1}] } {Tr [ T^{L}]}  ~;
  ~ \rho_0 = \langle  e_i \rangle= z  \frac{ Tr [ E T^{L-1}] } {Tr [ T^{L}]} ; \cr
  \rho &=& \langle a_i \rangle=  \frac{ Tr [ A T^{L-1}] } {Tr [ T^{L}]}.   
 \end{eqnarray}
Here  $e_i,a_i,d_i$  are  site  variables  which   are  unity   when  the  site $i$ is    
vacant, occupied   by an engine,  or by other  units  of a $k$-mer, respectively; otherwise, $e_i,a_i,d_i$ are  $0.$  
In the thermodynamic limit,
these   definitions of densities  are   equivalent to  Eq. (\ref{eq:dens-fug}).

\subsection{Correlation functions}
 Now, we proceed to calculate  the  two point correlation functions.  The  engine-engine correlation 
function is 
{\small
\begin{equation}
C(r)= \langle  a_i a_{i+r} \rangle   -  \langle  a_i  \rangle ^2 =  
 \frac{ Tr [ A T^{r-1} A T^{L-r-1}] } {Tr [ T^{L}]} - \rho^2, \label{eq:corr1}
 \end{equation}}
which,   in the thermodynamic limit, can generically  be expressed  as  {\small
\begin{equation}
C(r)=  p_1(z,x;v)  \left( \frac{ \lambda_1}{\lambda_{max}}\right)^r  + 
p_2(z,x;v)  \left( \frac{ \lambda_2}{\lambda_{max}} \right)^r,\label{eq:corr2}
\end{equation}}
where $p_1$ and $p_2$ are   independent of $r.$  The behavior  of the  correlation functions  depend  on the 
nature of eigenvalues $\lambda_{1,2},$  which can be   determined from the  properties of the 
characteristic  function $f(\lambda)$, in Eq. (\ref{eq:char}).     
 When   the    eigenvalues $\lambda_{1,2}$ are real,  ordered    as  $\lambda_{max}
>|\lambda_1|>|\lambda_2|,$  the system has  two  length scales $ |\ln \frac{|\lambda_{1,2}|}{\lambda_{max}}|^{-1}$ 
and  the asymptotic  form  of  the  correlation    function is  dominated   by   the largest  one, 
$\xi=|\ln \frac{ |\lambda_{1}|}{\lambda_{max}}|^{-1}.$  Correspondingly, the  correlation has a monotonic exponential 
decay
\begin{equation}
 C(r) \simeq p_1(z,x;v) e^{-r/\xi}.
\end{equation}

Also,  in some parameter zone eigenvalues  $\lambda_{1,2}$   may  become  complex. 
As they appear as complex conjugates  we write  $\lambda_{1,2} = \bar \lambda  e^{\pm i \theta},$ with 
$\bar \lambda < \lambda_{max}.$ In this regime, 
$p_{1,2}$  defined in Eq. (\ref{eq:corr2})  must be  complex conjugates 
$p_{1,2} = \bar p  e^{\pm i \phi}$ so that the correlation   function  $C(r)$  is  real.
Consequently   $C(r)$  shows  a damped oscillation, with  a generic  functional form 
\begin{equation}
 C(r) = \bar p   e^{-r/\xi}  \cos (\theta r + \phi),
\end{equation}
where  $\xi^{-1}=  |\ln \frac{ \bar \lambda}{\lambda_{max}}|.$

It is   interesting  to note that  a   damped oscillation   of  the radial distribution function is a  
typical feature of  a system  in liquid phase,  in contrast to  the exponential decay of the same  in  the vapor phase. 
Naively  one would think  that  such   a change  is a  direct consequence   of  an underlying  liquid -vapor phase transition. 
However,  this is not the scenario  in CRR model;  possibility of  a  phase  transition is    ruled  out here   
as the   largest eigenvalue   $\lambda_{max}$   is  non-degenerate for any $x,v,z$ and  the corresponding free 
energy  $\ln \lambda_{max}$ would not  be   non-analytic   
anywhere. This phenomena, a  macroscopic  change in the nature of correlation function  in absence of  any phase transition, 
has   been  known  for a while in literature, in different contexts,  under the name of  
{\it disorder  points} (or {\it lines)}.  In a broad  sense,  the {\it disorder points}  separate the regions 
in parameter space showing qualitatively different pair correlation functions and were  first  introduced by Stephenson 
\cite{step_1}. For Ising  chains with ferromagnetic nearest-neighbor  and anti-ferromagnetic next-nearest-neighbor 
interaction, the spin-spin correlation function  in the disordered paramagnetic phase $(T>T_c)$ shows damped oscillations   when  
$T$ is  larger than  $T_D$ (the disorder point) whereas  it decays exponentially   for $T_c<T<T_D$ \cite{step_2}.
There are several  other lattice spin models  \cite{aspects_dl, latticespin_dl} that  exhibit disorder lines. 
In context of fluids, a similar behavior  has been observed  in  the decay of density profiles  and the  radial 
distribution functions - here  the disorder lines  which separate different regimes  are  conventionally  
termed as Fisher-Widom  lines \cite{fisher,evans}. Recent studies   also indicate existence  of  disorder lines   in 
phenomenological models  of  QCD at finite temperature and density \cite{nishimura_2015}.

 \begin{figure}[t]\begin{center} 
 \includegraphics[width=8 cm ]{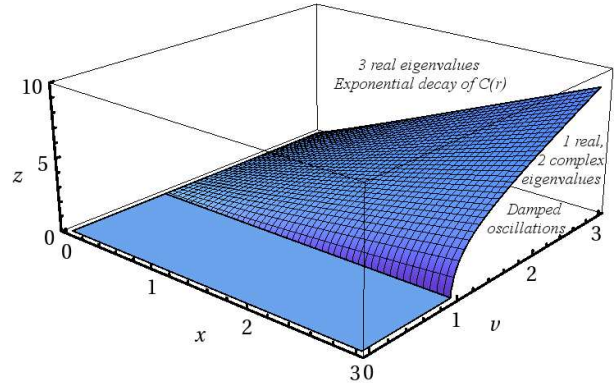} \vspace*{-.9 cm}
\caption{(Color online) Disorder surface separating regions where pair correlation function  
$C(r)$ decays exponentially (all three  eigenvalues are real) from   regions  having  damped  oscillations 
(where the  sub-dominant  eigenvalues are  complex).} 
\label{fig:roots}\end{center}
\end{figure}

In  CRR  model,  there  are three   parameters - the monomer diffusion rate $v$  and  the fugacities   $x,z$ 
(or alternatively   the   densities  $\rho,\eta$). 
Thus  we have a  two dimensional {\it disorder  surface}    that separates the  regimes of exponential decay from that of the damped oscillatory 
correlation functions in the $(x,v,z)$ space.
To identify  the disorder  surface  we    study   the  generic features of the   characteristic   function $\lambda f(\lambda)$ taking  help of the  Descartes' 
sign  rule.  
Since,  all the parameters  $x,v,z$ 
are positive,  Descartes'  sign  rule  indicate that  there is  exactly  one negative root  when  $z+v<1 $ irrespective of the value 
of $x$. In this regime,  the  root other  than $\lambda_{max}$   must be real, 
as  complex  roots are generated pairwise.
 In the   region  $z+v>1,$  we  have at most  three positive roots: one positive  and two complex {\it or}  all positive;   
 the  disorder surface    that separates  these   regimes   in $(x,v,z)$-space   is  shown in Fig. 
 \ref{fig:roots}. In the  region where  the  sub-dominant  eigenvalues are `complex', i.e.,  $\lambda_{1,2} = \bar \lambda  e^{\pm i \theta}$  with  $\bar \lambda < \lambda_{max},$   the  correlation  function    exhibits  damped  oscillations.

The  exact  analytic   expression of eigenvalues, densities and   the  two point  correlation  functions  are  rather  lengthy. 
For  the purpose of illustration, we  provide the details in next section, for  the  CRR model only  with  $v=2$,   
which   lead  to   both  decaying  and oscillating  correlation functions  in different  
density  regimes separated  by a  disorder line.  
\subsection{CRR model   with   $v=2.$}

In this   sub-section  we  focus   on the CRR model for a special case $v=2$.  We  have  already set  the  reconstitution 
rate $\w=1,$ thus   the    grand canonical partition function  depends  only   on two   parameters  $z$ and $x$ which 
fix the densities $\rho_0$ and $\rho_d.$  The packing  fraction,  which  is  defined as 
the  fraction of the lattice occupied by $k$-mers, is  now  
$\eta =  \rho + \rho_d=1-\rho_0.$  
For $v=2,$  the eigenvalues are given by,
\begin{eqnarray}
  &&\{ \lambda_{max}, \lambda_{1},\lambda_{2}\}  =   \{ g(1), g(\Omega), g(\Omega^2) \} 
\end{eqnarray}
where $g(y)  =    a + s y - \frac{p}{s} y^2$  and  $\Omega= e^{i\pi/3}.$  
The other parameters    are,  
$a=3z+2x+2$ , $p= 4 (1+x)^2 + 3z ( z-2x)$ and $ q=8(1+x)^3  + 9 z (z-2x)(x- 2 z) $ and $s^3=q+\frac{1}{2}\sqrt{4q^2-p^3}.$ 
The $k$-mer  number density  $\rho= 1-\rho_0 - \rho_d$   and the  packing fraction  $\eta = 1 -\rho_0 $ are
now calculated  from Eq.  (\ref{eq:dens-fug}), 
 \begin{eqnarray}
\rho  &=&  \frac{(2\lambda_{max}-z)(\lambda_{max}-z)(\lambda_{max}-x)}{\lambda_{max}(6\lambda^{2}_{max}
  -2\lambda_{max}(3z+2x+2)+z^2+2z+3xz)} \cr 
  \eta &=& 
 \frac{ \lambda_{max}(2\lambda_{max}-z)(\lambda_{max}-z) +xz} {\lambda_{max}(6\lambda^{2}_{max}  -2\lambda_{max}(3z+2x+2)+z^2+2z+3xz)}.\cr&&
 \label{eq:density_v2}
  \end{eqnarray}   

\begin{figure}[h] \begin{center}   \vspace*{1. cm}
 \includegraphics[width=7 cm]{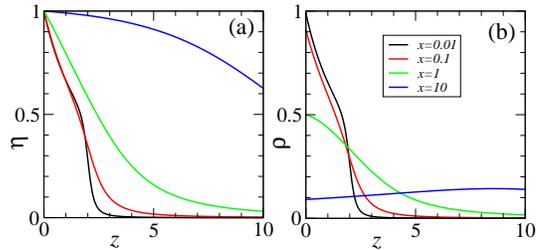}
\caption{(Color online)  Densities: (a)  Packing fraction $\eta$ and (b)  $k$-mer density  $\rho$, as a  function of  
$z$   for $x=0.01,0.1,1$ and $10.$}
\label{fig:density}\end{center}   
\end{figure}

In Fig. \ref{fig:density} we have   plotted $\eta$  and $\rho$ as a  function of $z,$ for different 
$x= 0.01, 0.1,1$ and $10$   respectively  in(a) and (b). 
As expected,  for $z =0,$    the packing  fraction is $\eta=1,$ independent of  the  fugacity $x.$   On the other hand, 
in the limit $x\to0,$ it appears that   both  $\eta$ and  $\rho$  might   become  discontinuous  at $z=2$  leading to a 
possibility of   phase  coexistence. In fact  {\it this} seems to be  the case for any $v>1$, and  we   discuss  this  possibility   separately  in the next section in details.   

\begin{figure}[h] \begin{center}   \vspace*{1. cm}
 \includegraphics[width=7 cm]{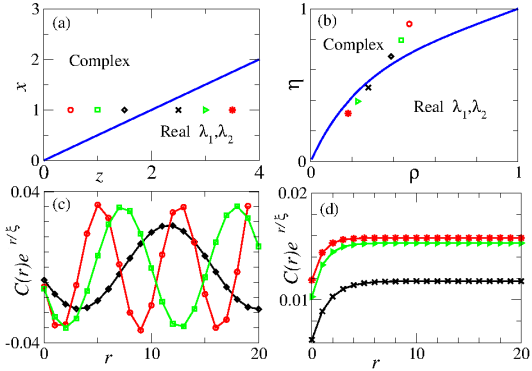}
\caption{(Color online) Correlation functions: (a) For  $v=2,$ complex roots  appear   for  $x>z/2$, (b)  the corresponding line   
in $\rho$-$\eta$ plane. (c) engine-engine correlation function $C(r)$  for $z= \frac{1}{2}, 1, \frac{3}{2} $ and 
(d)   $z= \frac{5}{2}, 3, \frac{7}{2}.$  For (c) and (d),  we use  $x=1$, and  the  symbols  used  here for 
different  $z$ are also marked  in  (a) and (b). }
\label{fig:corr}\end{center}   
\end{figure}

We now  proceed  to calculate the   two point   correlation functions,  first  the engine-engine correlation function $C(r)$
defined  in Eq. (\ref{eq:corr1}).  If we  formally write  the densities in Eq. (\ref{eq:density_v2}) as  functions of $z,x$ and $\lambda_{max},$  as  $\rho\equiv  \hat\rho(z,x;\lambda_{max})$  and  $\eta \equiv  \hat\eta(z,x;\lambda_{max}),$  
the   engine-engine correlation function, calculated  using Eq. (\ref{eq:corr1}),  can be written as
\begin{eqnarray}
 C(r)&=&\langle a_i a_{i+r} \rangle- \rho^2 =  \rho\left[ \hat \rho(z,x;\lambda_{1})
\left( \frac{\lambda_{1}}{\lambda_{max}}\right)^r \right. \cr&&   ~~~~~+\left.
\hat \rho(z,x;\lambda_{2})\left(\frac{\lambda_{2}}{\lambda_{max}}\right)^r \right].\label{eq:corr_v2}
\end{eqnarray}
Also, the   density  density correlation  function $\langle s_i s_{i+r} \rangle- \eta^2,$   takes a  form 
similar  to the   right hand side of the  above equation  with    $\rho \to \eta$ and   $\hat \rho(.) \to \hat \eta(.).$
 Clearly,   $C(r)$ will exhibit  damped oscillations  when  $\lambda_{1,2} = \bar 
\lambda  e^{\pm i \theta}$ are   complex.   Whether  such  a regime, separated  from  the  usual  exponentially decay  by 
a  disorder line,  exists for $v=2$, can be  determined from  the  characteristic  polynomial  
$\lambda f(\lambda).$ The  discriminant of the  $f(\lambda)$ vanishes  for  $x=z/2,$ which  is the    disorder 
line  for  $v=2$  (as  shown in Fig. \ref{fig:corr}(a)); correspondingly the disorder  line in 
$\eta$-$\rho$ plane   (shown in Fig. \ref{fig:corr}(b)) is 
 \begin{equation}
   \eta=\frac{\rho(\rho^2+3)}{(\rho+1)^2}
 \end{equation}
Thus,  for  $\eta > \frac{\rho(\rho^2+3)}{(\rho+1)^2}$   
the  correlation functions  are expected  to have   damped  oscillatory behavior    
whereas in the other regions  the  correlation functions must  decay exponentially as a function of $r.$ 
To illustrate this,  we   take $x=1,$   which  indicates that   the correlation functions  must be    oscillatory 
for any  $z<2.$   In  Fig.  \ref{fig:corr}(c) and (d) we  have   plotted  $C(r) e^{r/\xi}$   as a function of $r$ 
for $z= \frac{1}{2}, 1, \frac{3}{2} $  and  $z= \frac{5}{2}, 3, \frac{7}{2}$  respectively.     
All  these  $(z,x)$  values are  shown as symbols in Fig. \ref{fig:corr}(a). The corresponding densities 
are are  shown    in  $\eta$-$\rho$ plane,  marked  as same symbols (in Fig. \ref{fig:corr}(b)).  
Clearly $C(r) e^{r/\xi}$ shows  oscillations  when $z<2,$ whereas  it  asymptotically 
approaches to a  constant when  $z>2.$  It appears that  existence   of  disorder  lines 
is a  generic  feature in extended systems;  this may be  a consequence  of the   hardcore restriction among $k$-mers which 
mimics  a short range repulsion  existing in model fluids. \\

We close this section with  the  following interesting  remarks.\\

{\it Remark 1:} The  four dimensional  representation in  CRR model leads to a transfer matrix $T$ with $Det[T]=0.$
This in  turn  means  that one of the eigenvalues $\lambda=0,$  indicating  that  there  might  be   a three dimensional 
representation  which satisfy the matrix algebra  Eqs. (\ref{eq:main}) and (\ref{eq:de0}). 
We are able to  find one such representation, 
\begin{equation}
\begin{array}{ccc}
 A=\left( \begin{array}{ccc}
1 & 0 & 1 \\
1 & 0 & 1 \\
0 & 0 & 0 \\
\end{array}
\right), &
D=\left( \begin{array}{ccc}
0 & 0 & 0 \\
0 & 0 & 0 \\
0 & \w & \w \\
\end{array}
\right), & E=\left( \begin{array}{ccc}
 \frac{1}{v} & 0 & 0 \\
0 & 1 & 0 \\
0 & 0 & 0 \\ 
\end{array}
\right) \\
\end{array}
\end{equation}
It is   easy  to check that $T= A+x D+z E$   gives  the same  characteristic polynomial $f(\lambda)$ as in Eq. (\ref{eq:char}).

{\it Remark 2:}  For  CRR model with  $v=2,$  the line  $x=z/2,$ which  separates  regions  having  damped oscillations  
from regions with exponentially decaying correlations is very  special. On this line the  engine-engine   correlation function 
$C(r)$    vanishes,  whereas   the density-density correlation  function   remains finite.  This  can be verified  from  
directly calculating  the eigenvalues on this line, which are $\lambda_{max}= 1+ 2x$ and $\lambda_1=x=\lambda_2.$ 
 This  makes the co-efficients  $\hat \rho(z,x;\lambda_{1,2})=0$ in  Eq. (\ref {eq:corr_v2}).

{\it Remark 3:}   
This  formulation is   inadequate   to  calculate   spatial  correlation functions  of $k$-mer models when 
reconstitution is absent.   Naively one may think,   setting  $\w=0$   could  work.  
But $\w=0$   would   impose  a condition   $\langle \beta|D^kE^n|\alpha \rangle =0$ for  all  $k,n$,    
which   forces   the  weight  of every  configuration $Tr[...AD^kE^nA...]$  to vanish.  
In  fact  when $\w=0,$   diffusion  of $k$-mers (of different size)  is  the only dynamics  on the  lattice, 
which keeps the initial  ordering of  their  size  invariant. Now, the configuration  space 
has infinitely many   {\it disconnected ordering-conserving sectors},  and  one must write  partition  sums 
separately  for  each sector.

{\it Remark 4:}  In contrast  to   constant-rate reconstitution  model,  one may define   a  constant-rate  diffusion 
model considering   $u(k)= v,$ a constant. Now, let   us  consider  a reconstitution rate  
$\w(k,k') = 1 + (\w-1)\delta_{k,2}\delta_{k',1},$  where  dimers reconstitute
with monomers at rate $\w$ where as any other two $k$-mers reconstitute  with  unit rate.
In this  constant-rate  diffusion model we  have  a simple  two dimensional representation:
\begin{equation}
 \begin{array}{ccc}
A=\left( \begin{array}{cc}
1 & 1 \\
0 & 0 \\
\end{array}
\right), &
 D= \left( \begin{array}{cc}
0 & 0 \\
w & 1 \\
\end{array}
\right), &  E=\left( \begin{array}{cc}
\frac{1}{v} & 0 \\
0 & 0 \\
\end{array}
\right) \\
\end{array}.
\end{equation}
In a  two dimensional representation both  eigenvalues of  $T= A+xD +zE$ (which is a positive matrix)  are  real and 
hence   possibility  of oscillatory   feature  in spatial correlation functions  is ruled out.  
\subsection{ Phase coexistence  in  CRR model in $x\to 0$ limit}
In this section  we  investigate the CRR model in $x\to 0$ limit.   We  have seen  in the previous section that  for $v=2$,
the  $k$-mer density $\rho$  shows  a  sharp drop  at   $z=2.$  In fact this  feature is  quite  generic  for  all $v>1.$ 
When  the fugacity $x$ associated with $D$s   approaches  $0,$  we  
expect  a  microscopic number of  $D$s in the system. In other words  most  
$k$-mers are    only monomers. Thus, the best case scenario that represents  $x\to 0$ limit is 
a  system with {\it one}  single $D.$  
Since  this  $D$   must  be associated with an engine $A,$  
we have exactly  one   dimer  in the  system which diffuses  in the system with unit rate, 
and  all other  $k$ mers are monomers diffusing  with rate $v.$
Thus,  the   dimer can be  considered as a  defect particle  in  the system. 
The re-constitution  can  occur  only  between this  dimer and an adjacent    monomer (when both are immobile) 
with    rate $\w=1.$   In this  case, the reconstitution is  equivalent to  exchange  of a monomer  and a dimer.

Representing the  single dimer as $2$ and the monomers as $1$s,   and vacancies as $0$s, the   dynamics of the system  can be written as
\begin{equation} 
  10   \xrightarrow{v} 01 ~~;~~20   \xrightarrow{1} 02 ~~;~~ 211   \xrightleftharpoons[1]{1} 121. 
 \label{1dimerA}
\end{equation}
Since this  dynamics  is only a  special  case of the CRR  model, we can  proceed  with  the     
$4 \times 4$  representation given in Eq. \eqref{eq:rep_v2}. However,  in this  case 
there  is a valid $2$-dimensional representation,  because the weight of
every configuration of the  system   can  be  written  here  as  $Tr [ A D \prod_{j=1}^{L-2} X_i],$  
with   $X_i$   being  either  $E$  or $A.$  Clearly, these matrix  strings  do not  contain $ED$
and one  need  not  bother  about the constraint $ED=0$  in  Eq. (\ref{eq:de0})
and can work   with  the $2\times 2$ matrices  $\{\tilde A , \tilde D, \tilde E\}.$   
In other words,  the  dimer,  monomer and vacancy    are  now represented by   
$D_2= \tilde A \tilde D,$    $D_1= \tilde A,$  and  $\tilde E$  respectively.
In GCE,  the   partition function is now 
\begin{equation}
 Z  =  Tr  [ D_2  T^{L-2}]   ;  T = z \tilde  E  + D_1  =  \left( \begin{array}{cc}
1+z/v & 0 \\
1 & z \\
\end{array}
\right).
\end{equation}
The  eigenvalues of $T$  are    $\{z, 1+\frac{z}{v}\}$  and thus, for $v>1$  the maximum eigenvalue 
changes from  being $\lambda_{max} =  1+\frac{z}{v}$ to   $\lambda_{max} = z$ at  a critical fugacity 
$z_c=\frac{v}{v-1}.$  So, the   partition function,  in the  thermodynamic system, $Z \sim \lambda_{max}^{L}$ 
is  non-analytic at  $z= z_c$ indicating a   phase transition. 

First we   calculate the density profile, as  seen from the defect, which  
can be  expressed as 
\begin{equation}
\rho(r)=\frac{ Tr[D_2 T^r D_1 T^{L-3-r}] } {Tr[D_2 T^{L-2}]}= \frac{\gamma}{z} \frac{z \gamma^r -z_c}{z \gamma^{L-2} -z_c}
\end{equation}
where  $r$  is distance from the defect site (or  the dimer)   and $\gamma = \frac{vz}{v+z}$   the    
ratio  of  eigenvalues.  Thus, when   $z<z_c$ or  $\gamma <1,$  the   profile  $\rho(r)$   has a   boundary 
layer  in front of the defect site  which  extends  over a length scale  $1/|\ln\gamma|$  and   for 
large $r\ll L,$    it  saturates to a  value   $\rho_s=\gamma/z =  v/(v+z).$ 
Thus for a  thermodynamically large system,  the bulk  of the   system has a  density  $ v/(v+z),$   which  is  
same as  the  expected  monomer density   in the thermodynamic limit, 
$$\rho= 1- z\frac{d}{dz}\ln \lambda_{max} = \frac{v}{v+z}.$$

\begin{figure}[t]\vspace*{.3 cm}
 \centering \includegraphics[width=8cm]{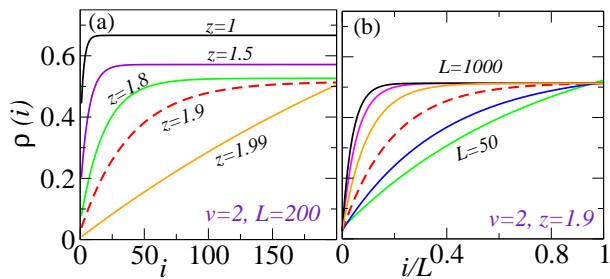}
\caption{(Color online) (a) The density profile  $\rho(i):$ (a)  for  $L=200, v=2$  and    different 
$z=1,1.5,1.8,1.9,1.99.$  For small $z,$  $\rho(i)$  saturates  to  the   macroscopic  
bulk density  $v/(v+z).$  For  large $z$ near $z_c$,  such a saturation can be  seen  by increasing 
the  system size $L$ - this is shown in (b).  The  profile   for $z=1.9$ (marked as   
dashed line)  as a function of $i/L$   shows  saturation  as  $L$ increased.  
}
\label{fig:prof}
\end{figure}
In Fig. \ref{fig:prof}(a)  we have  plotted   $\rho(r)$ for different $z$  considering  $v=2$  
(corresponding  bulk densities  are   $\rho_s= 2/(2+z).$  
It is evident  that   for a small system (here $L=200$)  the  boundary layer   
invades into the  bulk as $z$ approaches  the critical value $z_c.$  However,  for  any 
$z \lessapprox z_c,$   the boundary layer shrinks to $r=0$  in the  thermodynamic limit $L\to \infty.$ 
This  is shown in Fig. \ref{fig:prof}(b)  for $z=1.9$  ( here $z_c=2$)  and   
$L=50,100,200,400,700,1000.)$ 

The fugacity $z$ associated with  the   vacancies  can   tune the density  of the monomers in the   regime
$z<z_c,$  which  corresponds to a density regime $\rho_c<\rho<1,$ where $\rho_c= 1- 1/v.$ 
In  the  canonical  ensemble,  if the conserved density  is   fixed at  some  value  $\rho <\rho_c,$
the system is  expected  to  show  phase coexistence. Since the allowed    macroscopic densities  are 
$\rho_c$  and $0,$  and    the  average density  has to be $\rho$,    
the   system would  allow  a local density  $\rho_c$ in $\delta=\rho/\rho_c$ 
fraction  of the lattice  and keep $1-\delta$ fraction vacant.

The  single dimer  problem we discussed here  is  very similar to the   single defect  in totally asymmetric 
simple exclusion process  (TASEP) studied earlier in \cite{Mallick1996, Derrida1999}. 
This TASEP  model comprises of a single defect
particle (denoted  by $2$)  and $L-1$  normal particles ($1$s) on a ring
of size $L$, following a  hopping dynamics,
\begin{equation} 
 10   \xrightarrow{1} 01 ~~;~~20   \xrightarrow{\alpha} 02 ~~;~~ 21   \xrightarrow{\beta} 12.  \label{eq:defect}  
\end{equation}
A special case  of the  model $\alpha=1=\beta$ \cite{Lebowitz} corresponds to a scenario where  
the defect  $2$ is a  second-class particle   which  helps in locating the  shocks, if any. 
The model defined  by the  above dynamics, can be solved using   matrix product ansatz \cite{Mallick1996, evans_asep}, 
but the matrices $\{E,D,A\}$  corresponding to  $\{0,1,2\}$   have an infinite dimensional 
representation, closely related   to the    matrices  $\{E,D\}$ in  TASEP \cite{derrida_mpa}.
A  novel  feature that arises  in  this model   is the phase coexistence -  for $\beta <\rho <1- \alpha,$ 
the  system shows a  coexistence between a region of low-density   $\beta$  in front of the defect, 
 and a high-density $1 -\alpha$  behind it.  Thus a localized  shock    is  formed 
at  the site $j=\delta N,$ such that  the conserved density $\rho=  \beta \delta + (1-\alpha) (1-\delta).$  
In the following, we   compare  the phase coexistence  scenario of this  model  with  the single  dimer  
model studied here.

The dynamics  of the  single  dimer  model, Eq.  (\ref{1dimerA}), is equivalent  to 
\begin{equation} 
 10   \xrightarrow{1} 01 ~~;~~20   \xrightarrow{1/v}   02 ~~;~~ 211   \xrightleftharpoons[1/v]{1/v} 121.  
 \label{1dimerB}
\end{equation}
However, in contrast to   Eq. \eqref{eq:defect} there are two major  differences in (\ref{1dimerB}). 
Firstly the  dimer  occupies two lattice  sites, whereas the defect particle $2$ 
in exclusion   processes occupies only one lattice site - this difference is  not crucial in the  
thermodynamic limit. Secondly, in comparison to the    defect dynamics    in  Eq. (\ref{eq:defect}),  
(i) the dimer  can  exchange  with a  monomer in  both directions,   and (ii)   the exchange   
occurs {\it  only} when the immediate  right neighbor of the exchanging  sites  is occupied.

If we overlook these  differences,  the   models  are similar  when  $\alpha=  1/v = \beta.$
Thus  one   may speculate that a  phase   coexistence   may occur  when   $\frac{1}{v} <\rho < 1-   \frac{1}{v}.$
In reality, however,   for dynamics (\ref{1dimerA}),    phase   coexistence  occurs  when $0<\rho < \rho_c=1-   \frac{1}{v}.$ 
This  similarity is striking - particularly when  the  matrix  representation for  (\ref{eq:defect}) is infinite dimensional 
whereas the  same for (\ref{1dimerA}) 
is much  simpler, $2\times 2$ matrices.

\section{Conclusion \label{sec:conclude}}
  In this   article we   provide a  general formulation  to write the  steady state weights of  
  reconstituting  $k$-mer models  in  matrix product form. In the matrix formulation, we represent 
  a vacancy  by a  matrix $E$,  the   engine  of a 
  $k$-mer (leading  monomer unit from the left) by $A$ and rest of the monomer units  by $D$s. 
  In these  models,   the  $k$-mers, which  are extended  objects of different sizes,  move  to  neighboring  vacant sites  with  
  a  rate   that depends  on their  size. Reconstitution  can occur  among a  pair of $k$-mers,
  when they are in contact,   with a  rate  that depends  on   size of participating  objects.
  Reconstitution  usually means  transfer of a  monomer  from one  $k$-mer to  the other one; 
  we restrict   such a   transfer  if  length of   the  $k$-mer transferring  a monomer,  is unity. 
 This keeps the  number of $k$-mers conserved.    These  models can not be solved exactly for any arbitrary 
  diffusion  and  reconstitution rates. 
Some   of them   can be solved,  under quite general  conditions,  using a  mapping   of  $k$-mer 
models to a  two species  misanthrope process. These   exactly solvable   models, though simple, 
capture   the  different  phases  of the system and possible  transitions  among  
them \cite{daga_kmers,Daga2016} quite well. 
However, calculating  spatial correlation functions through  this mapping  is usually a formidable task, as  the mapping 
does not   keep track  of  the  site-indices and  the  notion of distance.  Thus, rewriting the  steady state weights 
in terms of a matrix product form  is greatly useful. 

If  $k$-mers  are  mono-dispersed, i.e.,  each  one  has a  fixed length $k$, the  matrix product  form is relatively 
simple \cite{gupta_kmer}. This is  because,   the  density   of $k$-mers   $\rho$  dictates the packing fraction  $\eta=\rho k$, 
and  one  can   get away with two matrices   $D$ and $E$ representing  whether the site is occupied or not.  The fact,     
poly-dispersed systems studied here have {\it two} independent conserved  quantities $\rho$ and $\eta$, brings in additional  
complications. First,  to write   the steady state  in  matrix    product form,   we need an additional matrix 
$A$ (along  with  $D,E$)  to  identify the  $k$-mers  and to keep track  of the  conserved $k$-mer  number.  
Next,  additional care  must be taken   to ensure that    every   block of   vacancies must  end  with $A$- in other words 
all configurations must  be devoid of  $ED.$

In summary, the reconstituting  $k$-mer models  for which  the   steady state  weights  can be obtained exactly  through a two species  
misanthrope process,  we  device  a  matrix  formulation to  calculate   spatial correlation functions explicitly. 
The  required  matrix algebra, and a generic representation    that satisfies this  algebra are also  provided.  
Specifically, we   demonstrate  the  formulation  for the constant-rate reconstitution (CRR) model where
reconstitution occurs with a constant rate $\w,$ and all  $k$-mers except the  monomer, move  to  its right neighbour (if vacant)
with unit  rate; the rate   for  monomers  is  $v\ne 1.$  The  two point  spatial correlation  functions  of  CRR model 
show interesting behavior when $k$-mer density $\rho$ and packing fraction  $\eta$ are tuned.  In some density regime  the spatial  correlation 
functions  show damped  oscillation whereas    in other regimes they   decay   exponentially.  The  boundary that  separates 
these regimes  in   $\rho$-$\eta$  plane, conventionally  known  as
the  {\it disorder  line,} is  calculated    analytically. 

A special limit $\eta\to \rho$  of the  CRR  model  is  best represented  by  a  system  of  monomers and  a {\it single} dimer. 
The reconstitution process  in  this  single dimer model is  equivalent  to  exchange   of  a monomer   with  the  dimer, when 
in contact.    Effectively, the dimer   behaves like a  defect  in the system  and  exhibits a  phase  coexistence, similar to   
the  one observed in   asymmetric exclusion processes with a  single defect.  
We  must mention that,  though   both  models capture   the   phase separation scenario,    the   single dimer  model    
is  represented by   a set of  simpler  $2\times 2$ matrices in contrast   to  the  infinite dimensional representation in  exclusion  processes with a defect.

In all through this article we  have considered  directed diffusion of $k$-mers. This provides a  natural interpretation  that 
reconstitution   occurs   among  immobile  $k$-mers.    However, the steady state   and  thus  physical properties of the model  are   
invariant if we use,   a  symmetric  diffusion of   $k$-mers,  and a  reconstitution process   that 
does not allow  two particles  of  different species  (in  corresponding two species misanthrope process)   
to  move out of a   box simultaneously.

In  this  article   we   provide  matrix  product  steady states   for   a  class   of   $k$- mer  models, by mapping   them to 
a   two species    misanthrope process. In fact   any   two species    misanthrope process 
can   be  mapped  to  a  lattice  containing  extended objects -  the  steady state 
of  such systems   can {\it always} be  be written  in   matrix product  form. 
We  believe  that 
the matrix  formulation developed here  can  be  useful,  in  general,  to   explore    spatial correlation 
functions  in  extended systems  in one dimension.


\begin{thebibliography}{99}
\bibitem{footnote_TMAP}  In  misanthrope process \cite{EvansBeyondZRP}, a particle   hops  from a site to one of its 
 neigbours with a  rate  that depends  on the   number of particles  of  both  the  departure and 
the arrival sites. A two   species misanthrope process (TMAP) is a    generalization, where the hop rate of each  
species   depends  on  occupation number of  both  species    in   departure and arrival sites.
\bibitem{footnote3}  Additional care 
must be taken  at  the transition  point \cite{Kavita_Arvind}, where   ensemble equivalence   is in question.
\end{thebibliography}
\end{document}